\documentclass[twocolumn, superscriptaddress, prl]{revtex4-1}
\usepackage[total={18.2cm,24.4cm},centering]{geometry} 

\usepackage{times}
\usepackage{graphicx}	

\usepackage{color}

\usepackage{soul} 
\usepackage{amsmath}
\usepackage{mathrsfs}
\usepackage{hyperref}
\usepackage[normalem]{ulem}
\usepackage{physics}

\hypersetup{colorlinks=true, citecolor=cyan, urlcolor=blue, linkcolor=blue}
\DeclareMathOperator{\sinc}{sinc}

\newcommand*\subtxt[1]{_{\textnormal{#1}}}
\DeclareRobustCommand\_{\ifmmode\expandafter\subtxt\else\textunderscore\fi}

\usepackage{newfloat}
\DeclareFloatingEnvironment[name={Extended Data Fig}]{extfigure}

\begin{document}

\title{A photonic quantum engine driven by superradiance}
\author{Jinuk Kim}
\author{Seung-hoon Oh}
\affiliation{Department of Physics and Astronomy \& Institute of Applied Physics, Seoul National University, Seoul 08826, Korea}
\author{Daeho Yang}
\affiliation{Samsung Advanced Institute of Technology, Samsung Electronics, Suwon 16678, Korea}
\author{Junki Kim}
\affiliation{SKKU Advanced Institute of Nano Technology, Sungkyunkwan University, Suwon 16419, Korea}
\author{Moonjoo Lee}
\affiliation{Department of Electrical Engineering, Pohang University of Science and Technology, Pohang 37673, Korea}
\author{Kyungwon An}
\email{Correspondence to: kwan@phya.snu.ac.kr}
\affiliation{Department of Physics and Astronomy \& Institute of Applied Physics, Seoul National University, Seoul 08826, Korea}

\begin{abstract}
\end{abstract}
\maketitle

\begin{bf}
Performance of nano- and micro-scale heat engines can be improved with a help from quantum mechanical phenomena\cite{scully2011quantum, wang2012efficiency,huang2012effects,rossnagel2014nanoscale,gelbwaser2015power,liu2016maximum,daug2016multiatom,yin2017optimal,klaers2017squeezed,deng2018superadiabatic,peterson2019experimental}.
Recently, heat reservoirs with quantum coherence have been proposed to enhance engine performance beyond the Carnot limit even with a single reservoir\cite{scully2003extracting,hardal2015superradiant}. However,
no physical realizations have been achieved so far.
Here, we report the first proof-of-principle experimental demonstration of a photonic quantum engine driven by superradiance employing a single heat reservoir composed of atoms and photonic vacuum. 
Reservoir atoms prepared in a quantum coherent superposition state underwent superradiance while
traversing the cavity.
This led to about 40-fold increase of the effective engine temperature, resulting in a near-unity engine efficiency.  Moreover, the observed engine output power grew quadratically with respect to the atomic injection rate.
Our work can be utilized in quantum mechanical heat transfer\cite{rodriguez2013thermodynamics,ronzani2018tunable} as well as in boosting engine powers, opening a pathway to development of photomechanical devices that run on quantum coherence embedded in heat baths. 
\end{bf} \\

Quantum heat engines, first considered by Scovil and Schulz-DuBois in 1959\cite{scovil1959three}, can outperform classical counterparts by utilizing quantum mechanical principles. They also serve as platforms for studying thermodynamics in the quantum regime\cite{allahverdyan2008work,zou2017quantum,ma2018universal, klatzow2019experimental,von2019spin,van2020single,bouton2021quantum}. 
A quantum engine generally operates between non-thermal and thermal reservoirs. Numerous proposals have been made for non-thermal reservoirs based on superposed\cite{scully2003extracting}, squeezed\cite{huang2012effects, rossnagel2014nanoscale} and entangled\cite{daug2016multiatom} states.
Experimentally, quantum engine cycles have been implemented in various systems such as cold atoms\cite{zou2017quantum,bouton2021quantum}, a nanobeam\cite{klaers2017squeezed}, an ensemble of nitrogen-vacancy centers\cite{klatzow2019experimental}, nuclear spins\cite{peterson2019experimental} and trapped ions\cite{von2019spin,van2020single}.

One of the unusual properties of the quantum engine is that the effective temperature of the engine can be higher than that of the reservoir in the steady state.
It was pointed out that such unbalanced temperatures enable an efficiency beyond the Carnot efficiency, a classical limit determined by the ratio of reservoir temperatures\cite{scully2003extracting,rossnagel2014nanoscale}. High efficiency, however, often accompanies low output power. It is thus of practical importance to seek for power enhancement as well, in addition to pursuing high efficiency. In this regard, enhancement by noise-induced coherence\cite{scully2011quantum} and correlated thermalization\cite{gelbwaser2015power} have been investigated theoretically. Increasing the efficiency at a maximal output, known as the Curzon-Ahlborn efficiency, has been proposed\cite{allahverdyan2008work, wang2012efficiency, rossnagel2014nanoscale} and demonstrated\cite{klaers2017squeezed}.

A promising -- but not yet implemented -- approach for achieving the aforementioned goals is to utilize 
a heat reservoir with quantum coherence among energy levels of constituent atoms\cite{scully2003extracting} or
a superradiant matter as a non-thermal reservoir of the quantum engine\cite{hardal2015superradiant}. 
In superradiance, emitters collectively interact with light, resulting in enhanced radiation\cite{yamamoto1999mesoscopic}.
The most representative characteristic of superradiance is that the emitted power is proportional to the square of the number of emitters. It was proposed to exploit this nonlinearity in photonic engines to enhance the engine output power greatly\cite{hardal2015superradiant}.

In this paper, we report  the first proof-of-principle experimental realization of a superradiant photonic engine employing a heat reservoir composed of atoms and a photonic vacuum. 
The engine's working fluid is a {\em photon} gas exerting a pressure on cavity mirrors. 
Our engine utilizes a quantum cycle resembling the classical Stirling cycle and works between a superradiant state and an effective thermal state of atoms with both at the same reservoir temperature. 
Owing to the non-passivity of the superradiant reservoir atoms, energy transfer from the reservoir to the engine was much enhanced\cite{niedenzu2016operation,francica2020quantum}.
This effectively raised the engine temperature by a factor of 40 over the thermal-state case, resulting in a near-unity engine efficiency up to 98\%. 
The engine power was also greatly increased, proportional to the square of the atomic injection rate, owing to its superradiant nature. 

\begin{figure*}
\centering
\includegraphics[width=0.95\textwidth]{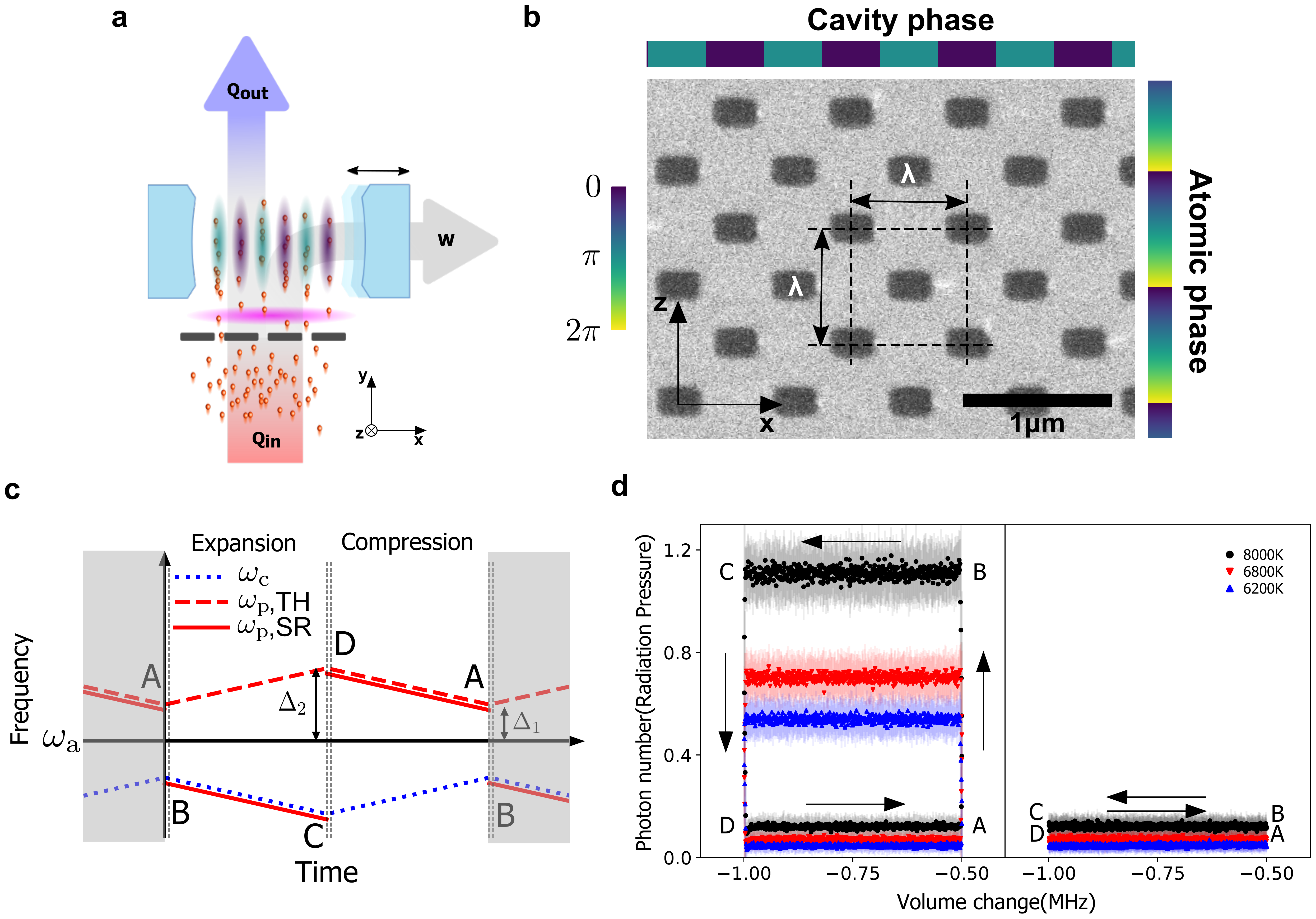}
\caption{{\bf Engine cycle and the pressure-volume diagram of the superradiant quantum engine.} {\bf a,} Schematic of the superradiant quantum engine. The mirrors of a Fabry-P\'{e}rot cavity(in light blue) act as a piston of an engine. Its working fluid is a photon gas exerting a radiation pressure on the mirrors. Atoms excited by a pump laser(in magenta) provide energy to the cavity. The injection location of atoms is set at the antinodes of the cavity mode(in blue chill and persian indigo) by a nanohole-array aperture(in dark grey). {\bf b,} Scanning electron microscope image of the nanohole-array aperture. The size of the rectangular hole is $0.35\lambda \times 0.25\lambda$, where $\lambda\equiv 791.3$nm. The distance between adjacent holes is equal to $\lambda$. {\bf c,} Pump and cavity frequency control during the cycle. Blue dotted and red solid(dashed) lines correspond to the cavity and pump laser frequencies of superradiant(thermal) photonic engine with abbreviation SR(TH), respectively. $\Delta_{1(2)}/2\pi\equiv0.5(1.0)$MHz. {\bf d,} The observed pressure-volume diagram of the engine cycle. The radiation pressure is represented by the photon number. The $x$ axis is the cavity-atom detuning, which is associated with the cavity mode volume (see Supplementary Note 6 for details). Black circles, red squares and blue triangles correspond to the cases when the effective temperature $T\_R$ of the reservoir is 8000K, 6800K and 6200K, respectively.  Experiments were performed under the conditions $g\tau=0.17$ and $\bar{N}=0.8$, where $g$ is the atom-cavity coupling constant, $\tau$ is the transit time of the atoms across the cavity, and $\bar{N}$ is the mean atom number. The left(right) figure represents the case when the reservoir is in the superradiant(effective thermal) state during the expansion stage, B$\rightarrow$C. Error bars shown as shaded vertical lines are one standard deviations from repeated measurements.
}
\label{fig:1}      
\end{figure*}


Main components of our engine are a high finesse cavity, two-level atoms traversing the cavity and a nanohole-array aperture with the hole spacing matching the atomic transition wavelength as shown in Fig.~\ref{fig:1}a. 
Prior to the cavity, atoms are sent through the nanohole array aperture(Fig.~\ref{fig:1}b) and then excited to a superposition state with a pump laser. The role of the nanohole-array is to localize atomic positions in order to make the relative phase between the atomic state and the cavity field the same for all atoms  (see Methods for details).
The atomic state is then given by
\begin{equation}
\left|\Psi\_{\rm atom}\right\rangle=\prod_{k=1}^{N}\left[\sin(\theta/2)\left|{\rm g}\right\rangle_k +\cos(\theta/2) e^{-i(\omega\_p-\omega\_c) t_k} \left|{\rm e}\right\rangle_k \right]
\label{eq:1}
\end{equation}
in the rotating frame of the cavity field, where $N$ is the number of atoms in the cavity, g(e) represents the ground(excited) state of the atom, $\omega\_p$($\omega\_c$) is the (resonance) frequency of the pump laser(cavity), $t_k$ is the arrival time of the $k$th atom at the cavity and $\theta$ can be viewed as the polar angle in the Bloch sphere.
When the pump and the cavity frequencies are equal, all atoms have a common phase, so the atomic state is an atomic coherent state or a Bloch state\cite{yamamoto1999mesoscopic,friedberg2007dicke}, 
$\left|\Psi\_{atom}\right\rangle=\prod_{k=1}^{N} \left[\sin(\theta/2)\left|{\rm g}\right\rangle_k +\cos(\theta/2)\left|{\rm e}\right\rangle_k \right]$.

The atomic coherent state exhibits superradiance \cite{le1993generation,yamamoto1999mesoscopic,friedberg2007dicke} as verified recently in experiments\cite{kim2018coherent}. 
In this superradiance, coherence is injected by pump-laser phase imprinting and stored in the cavity\cite{le1993generation}.
Superradiance can take place when the relative phase between the injected atomic state and the cavity is constant, {\it i.e.}, the pump laser and the cavity are resonant. 
As long as the cavity retains the phase coherence of preceding atoms or, in other words, when the mean number $N\_c$ of atoms injected during the cavity-field decay time is larger than unity, 
superradiance occurs with the radiation strength proportional to $N\_c^2$\cite{kim2018coherent}.

If the pump laser frequency is far detuned from the cavity resonance, due to the random arrival time $t_k$ in the phase $(\omega\_p-\omega\_c)t_k$, the state of atoms is reduced to a thermal state  (see Methods for details). It is known that the atomic beam in this case acts as a thermal reservoir to the cavity in the Markovian regime $(g\tau\ll1)$\cite{steck2007quantum,daug2016multiatom,turkpencce2019tailoring}, which holds in our experiment, where $g$ is the atom-cavity coupling constant and $\tau$ is the transit time of the atoms across the cavity.
Therefore, the atoms can act as not only a thermal but also a coherence reservoir depending on their phases\cite{le1993generation,hardal2015superradiant,ma2021works}.

\begin{figure*}[t]
\centering
\includegraphics[width=\textwidth]{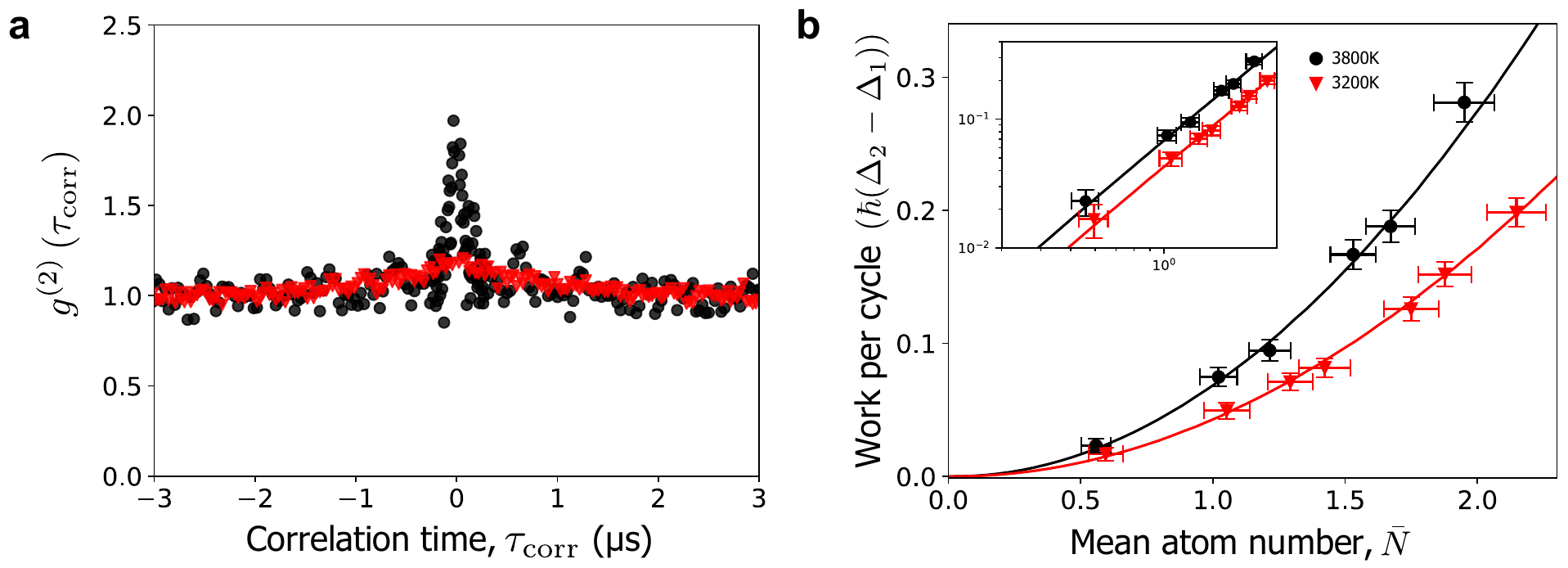}
\caption{{\bf The second-order correlation and the enhanced engine power by superradiance.} 
{\bf a,} The observed second-order correlation $g^{(2)}(\tau\_{corr})$ of the cavity field. Red triangles(black circles) correspond to $g^{(2)}(\tau\_{corr})$ in the expansion(compression) stage with the superradiant(effective thermal) state for the reservoir with $g^{(2)}(0)=1.20\pm0.01(1.97\pm0.08)$. {\bf b,} Dependence of the work done per cycle on the mean atom number for $T\_R=$3800K and 3200K, respectively. Black and red solid lines are numerical solutions obtained by the quantum trajectory simulation. The inset is a plot in the log-log scale. Fitted slopes are $1.91 \pm 0.04$ and $1.97 \pm 0.06$ for $T\_R$=3800K and 3200K, respectively. Experiments were performed under the  condition $g\tau=0.03$. Error bars shown are one standard deviations from repeated measurements.
}
\label{fig:2}      
\end{figure*}

Two highly reflective mirrors of the Fabry-P\'{e}rot resonator are subject to the radiation pressure of cavity photons and thus 
the mirrors can be viewed as a piston of an engine whose working fluid is a photon gas\cite{scully2003extracting,daug2016multiatom,turkpencce2019tailoring}.
The reservoir consists of an atomic beam and a photonic vacuum associated with the cavity decay (see Methods for details).
Our engine undergoes a four-stroke cycle and its timing sequence of engine control and the pressure-volume diagram are shown in Figs.~1c and 1d, respectively. 
Note that the resonance frequency of the cavity is modulated externally (see Supplementary Note 5 for details).
First, the cavity and the pump laser are set on resonance while the cavity volume is fixed (isochoric process A$\rightarrow$B).
The atoms are put in a superradiant state and the radiation pressure rises rapidly due to superradiance.
When the superradiance occurs, the steady state of the cavity photons becomes a thermal coherent state, {\em i.e.} a coherent state obtained by applying the displacement operator to a thermal state\cite{hardal2015superradiant} (see Supplementary Note 2 for details).
The system reaches the steady state within 1$\mu$s.
The observed second-order correlation function of the thermal coherent state is shown in Fig.~\ref{fig:2}a (in red dots).

In the expansion stage (B$\rightarrow$C), the resonance frequency of the cavity decreases as the cavity volume increases. To maintain the superradiance, the frequency of the pump laser follows that of the cavity on resonance.
After that, the pump laser frequency is instantaneously flipped with respect to the atomic resonance (C$\rightarrow$D), in order to disable the superradiance while retaining the effective temperature of the reservoir. 
With non-resonant pump and cavity, the radiation pressure drops rapidly to the same level as state A
due to the randomized phase of the atomic state.
In the compression stage (D$\rightarrow$A), the steady state of the generated photons is a thermal state as
verified by the second-order-correlation measurement in Fig.~\ref{fig:2}a (in black dots).
The process D$\rightarrow$A(B$\rightarrow$C) is an isoenergetic process\cite{liu2016maximum}, {\em i.e.}, the internal energy of cavity photons is constant, while the effective temperature of the reservoir remaining approximately constant (see Supplementary Note 5 for details). 

During the isochoric process A$\rightarrow$B with the cavity volume fixed, the internal energy of the engine or the number of photons increases while both the number of atoms in the cavity and the effective temperature of the reservoir are kept constant.
Introduction of the non-passive superradiant state in the reservoir in this process enables ergotropy transfer from the reservoir to the engine\cite{niedenzu2016operation,francica2020quantum}.
Ergotropy is the amount of energy that can be extracted from a non-passive state via unitary operations. 
Our engine is based on ergotropy, and more specifically it uses coherence.
Such coherence-induced ergotropy transfer can be explained in terms of a relative entropy of coherence for the reservoir\cite{ma2021works} (see Methods as well as Supplementary Note 4 for details).
On the other hand,
if the pump laser is detuned from the cavity resonance during the entire cycle, the cavity photons remain in a thermal state throughout.
In such a thermal heat engine, no work can be extracted from a single heat reservoir, as shown in the right figure of Fig.~\ref{fig:1}d.

Since the work is produced by the radiation pressure of the stored photons, it can be written as 
\begin{equation}
W=- n h \Delta \nu,
\end{equation}
where $n$ is the number of photons, $h$ the Planck constant and $\Delta \nu$ the cavity frequency difference associated with the volume change. The sign of $\Delta \nu$ is negative(positive) for expansion(contraction).
The work per cycle is then proportional to the photon number difference between the expansion and compression stages.
The thermal photon number $\bar{n}\_{th}$ grows linearly with the atomic injection rate while the photon number $\bar{n}\_{sr}$ by superradiance exhibits a quadratic growth with a small linear term associated with $\bar{n}\_{th}$. 
Therefore, the work per cycle is in proportion to the square of the atomic injection rate ($\gamma\_{inj}=\bar{N}/\tau$) or equivalently to the square of the mean atom number $\bar{N}$ as shown in Fig.~\ref{fig:2}b. 
The work per cycle for the engine cycle shown in Fig.~\ref{fig:1}d is 1.6, 2.1, $3.3\times10^{-28}$J at $T\_R=$ 6200K, 6800K, 8000K, respectively, where $T\_R$ is the reservoir temperature given by  (see Methods for details)
\begin{equation}
T\_R=\hbar \omega\_a\left\{k\_B\log \left[ \frac{\rho\_{gg}}{\rho\_{ee}}+\frac{2\kappa}{\gamma\_{inj}\rho\_{ee}(g\tau)^2 \sinc^2(\Delta\_{ac}\tau/2)} \right]\right\}^{-1},
\end{equation}
with $\hbar=h/2\pi$, $\omega\_a$ the resonance frequency of the atom, $k\_B$ the Boltzmann constant, $\rho\_{gg}$ and $\rho\_{ee}$ the diagonal elements of the density matrix of the injected atom, 
$\Delta\_{ac}$ the atom-cavity detuning and $\kappa$ the cavity decay rate.

The cavity field or the engine in the expansion stage can be regarded as being effectively hotter than the reservoir due to quantum coherence\cite{scully2003extracting,hardal2015superradiant}.
During the isochoric heating/cooling stage in Fig.~\ref{fig:1}d, there is no entropy change in the engine because the entropies of the thermal state and the thermal coherent state are the same for a given $T\_R$ (see Supplementary Note 3 for details).
It is the change in the reservoir's relative entropy of coherence that induces the energy transfer (or equivalently ergotropy transfer) from the reservoir to the engine or vice versa\cite{ma2021works} (see Supplementary Note 4 for details).
The engine entropy increases(decreases) in the expansion(contraction) stage, but the entropy change of the thermal coherent (or thermal) state in the expansion stage exactly cancels out that of the thermal state in the contraction stage, 
and therefore, the changes in the engine entropy are summed to zero around a complete cycle.

\begin{figure*}
\centering
\includegraphics[width=\textwidth]{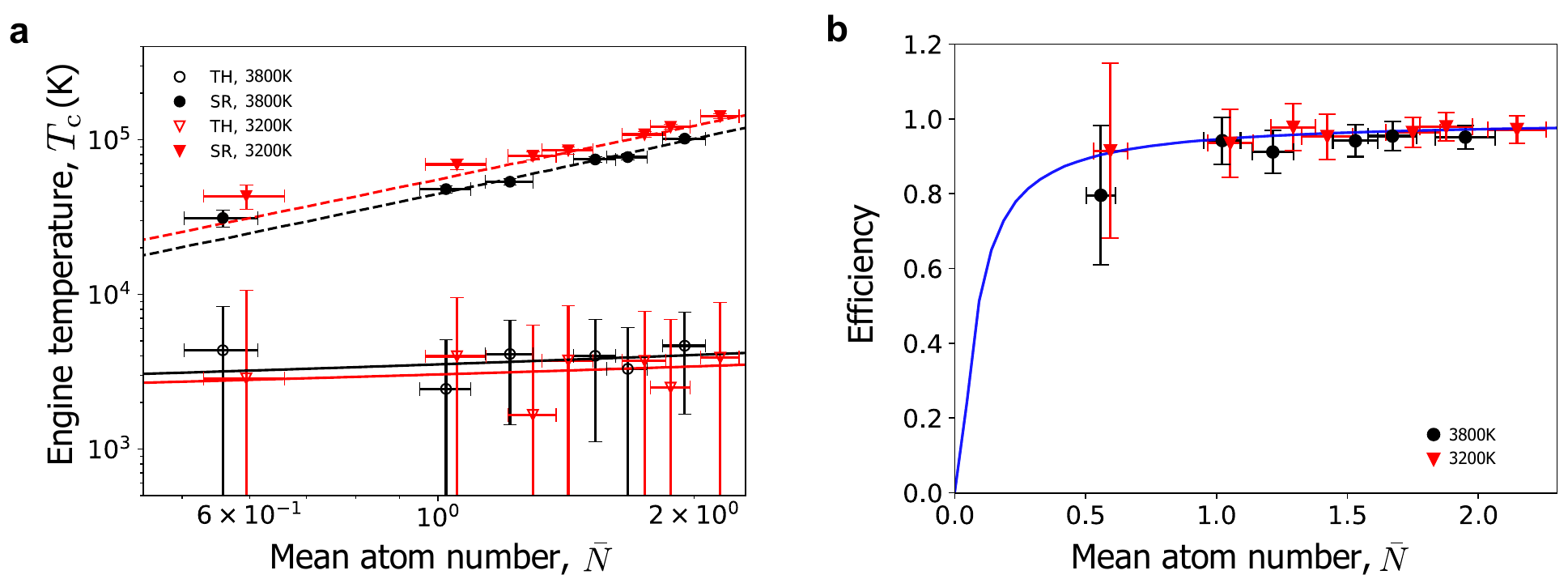}
\caption{{\bf Effective engine temperature and observed engine efficiency.} {\bf a,} Open(closed) symbols correspond the temperature of photons when the reservoir is the effective thermal[TH] (superradiant[SR]) state. Solid(dashed) lines represent the averaged reservoir temperatures(theoretical effective temperature of the photons). {\bf b,} Efficiency of the superradiant engine. The solid line is the numerical solution obtained by the quantum trajectory simulation. 
Experiments were performed under the condition $g\tau=0.03$. Error bars shown are one standard deviations from repeated measurements.
}
\label{fig:3}
\end{figure*}

During the expansion/contraction stage, the effective temperature $T\_c$ of the cavity photons or the engine, can be deduced from the thermodynamic relation $dQ=TdS$, among heat($Q$), temperature($T$) and entropy($S$), as
\begin{equation}
T\_c=\frac{n\hbar \omega\_c}{k\_B\log \left(1+\bar{n}\_{th}^{-1}\right) \bar{n}\_{th}}
\label{eq:2}
\end{equation}
(see Methods for details).
Equation~(\ref{eq:2}) applies to the case of superradiance with $n=\bar{n}\_{sr}$. For a thermal state $(n=\bar{n}\_{th})$, the photons are in equilibrium with the reservoir, and thus $T\_c$ equals $T\_R$ 
and becomes equivalent to the temperature of the Bose-Einstein distribution $ \bar{n}\_{th}=\left(e^{\hbar\omega\_c/k\_BT\_R}-1\right)^{-1}$.
The temperature $T\_c$ depends on the ratio of photon numbers $n/\bar{n}\_{th}$, which is proportional to the number of the atoms in the case of superradiance.
However, the meaning of the temperature given by Eq. (\ref{eq:2}) is different for the two cases.
For the thermal state, it is the usual temperature.
For the thermal coherent state, on the other hand, it is not the usual temperature because the state of photons is non-passive.
In that case, the effective temperature we have defined represents the temperature of an equivalent thermal state which would exhibit the same amount of energy exchange as the superradiant state.
Figure~\ref{fig:3}a shows the observed effective temperature of the engine or the cavity photons. We observe that $T\_{c,sr}/T\_{c,th}$ amounts to about 40 when $\bar{N}=2.1$ and $T\_R=$3200K (corresponding to the rightmost data points in Fig.~\ref{fig:3}a), where $T\_{c,sr}(T\_{c,th})$ denotes the effective photon temperature when superradiance occurs(when atoms are effectively in a thermal state). 
The effective engine temperature reaches up to $1.5\times10^5$K, which would be difficult to achieve with the conventional thermal heating due to heat loss to the environment, particularly for micro- and nano-scale engines with large surface-to-volume ratios.

In the thermodynamic limit, $k\_B T\_R \gg \hbar \omega\_c (\simeq \hbar \omega\_a)$ and $N \gg 1$, the isoenergetic process in Fig.~\ref{fig:1}d converges to the isothermal process\cite{liu2016maximum}, and therefore, our engine cycle approaches the classical Stirling cycle.
In our superradiant engine, the ergotropy input in the form of photons during A$\rightarrow$B is returned to the reservoir during C$\rightarrow$D.
So, the engine efficiency $\eta$ is determined solely by the ratio of the number of photons in the expansion and compression stages as (see Methods for details)
\begin{equation}
\eta=1-\frac{\bar{n}\_{th}}{\bar{n}\_{sr}}=1-\frac{T\_{c,th}}{T\_{c,sr}}.
\label{eq:7}
\end{equation}
The observed efficiency based on the experimentally observed photon numbers in the expansion and compression stages approaches unity as $N$ increases for a fixed $T\_R$ as shown in Fig.~\ref{fig:3}b, where up to $\eta=0.98\pm0.04$ is achieved. It is the highest ever recorded in quantum engines so far.

For a quantum engine operating with a small number of atoms, fluctuations in the atom number can affect the engine performance. In our superradiant engine, however, the engine performance is not degraded since the important atom number is $N\_c$, the number of atoms traversing the cavity during the cavity field decay time, which is given by $N\_c=\bar{N}/(\kappa\tau)\simeq 20\bar{N}\sim 40$, substantially large in Figs.~1d, 2b, 3a and 3b. 
The error bars there include fluctuations in $N\_c$ with the actual error further reduced by averaging the results of about 60 repeated measurements. 
In addition,
thermal excitation of phonons in the cavity walls can also affect the engine performance since it corresponds to an entropy flow without work. In our engine, the 
optomechanical coupling rate between phonons in the cavity walls and photons in the cavity is estimated to be about 40 mHz, so one can safely neglect the phonon excitation due to radiation-pressure changes.


To the best of our knowledge, our engine is the first experimental-realized quantum engine utilizing quantum coherence embedded in a heat reservoir. The quantum coherence in the reservoir made it possible to achieve
a near-unity efficiency as well as a greatly enhanced work scaling nonlinearly to the size of a thermal contact (intracavity atoms in our case) of a heat reservoir. 
The present work suggests an effective mechanism to implement quantum-enhanced heat transport\cite{rodriguez2013thermodynamics,ronzani2018tunable} and to boost the engine output by quantum coherence, opening a pathway toward photomechanical devices that run on quantum coherence.
Our work can further be developed to a versatile platform for fundamental studies on quantum phenomena in thermodynamics settings.

\section{Methods}

\noindent
{\bf Experimental setup.}
Our experimental setup employs an \textsuperscript{138}Ba atomic beam and a high finesse optical cavity.
The \textsuperscript{1}S\textsubscript{0}$\leftrightarrow$\textsuperscript{3}P\textsubscript{1} atomic transition at 791 nm is used. The cavity has the same resonance wavelength.
Experimental parameters are $(g,\kappa,\gamma)/2\pi=(334,74,25)$kHz, with the atom-cavity coupling constant $g$, the cavity decay rate $\kappa$ and the atomic decay rate $\gamma$ (all half widths). 
Here the quoted value of $g$ is a maximum value and it can be lowered by adjusting the atomic injection location across the cavity mode ($z$ direction in Fig.~\ref{fig:1}a).
The mean number of photons in the cavity, the mean number of atoms in the cavity, the distance between the mirrors (or the cavity length), and the second-order photon correlation in Fig.~\ref{fig:2}a are all experimentally measured quantities. These quantities are obtained, respectively, from the count rate of photons leaking out the cavity, the fluorescence from the atoms in the cavity, the resonance frequency of the cavity and the arrival times of photons leaking out the cavity registered on two single-photon counting modules in the Hanbury-Brown-Twiss arrangement.

Prior to the cavity and the pump laser, the atoms are sent through a nanohole-array aperture of checkerboard pattern (Fig.~\ref{fig:1}b).
More details on the nanohole-array aperture are described in Refs. \cite{lee2014three,yang2021realization}.
The distance between adjacent holes is the same as the resonance wavelength.
The phase of the cavity mode repeats alternatively between 0 and $\pi$ along the cavity axis due to the standing cavity mode structure.
On the other hand, the phase of the superposition state of injected atoms is imprinted by a traveling pump laser and it changes continuously along $z$ axis as shown in Fig.~\ref{fig:1}b. 
The role of the nanohole-array aperture is to pick up a common phase as well as the same atom-cavity coupling constant for all passing atoms.
Owing to the nanohole pattern, the relative phase between the atomic state and cavity field is constant.
Then the state of atoms can then be written as Eq.~(\ref{eq:1}) in the rotating frame at the cavity resonance frequency $\omega\_c$ with a phase factor $(\omega\_p-\omega\_c)t_k$ in the exponent. 
When the pump and the cavity frequencies are equal, this phase factor disappears and the atomic state becomes an atomic coherent state.
If the pump laser frequency is far detuned from the cavity resonance, atoms pick up the phase $(\omega\_p-\omega\_c)t_k$ at the time of arrival $t_k$ which is purely random  
since the time interval between two successive atoms obeys the Poisson statistics.
Due to the randomized phase factor, the state of atoms is reduced to an incoherent (thermal) state
\begin{equation}
\sin^2(\theta/2)\left|{\rm g}\right\rangle\left\langle {\rm g}\right| + \cos^2(\theta/2) \left|{\rm e}\right\rangle\left\langle {\rm e}\right|.
\label{eq:5}
\end{equation}\\

\noindent
{\bf Superradiance by coherence injection in a single reservoir.}
Time evolution of the cavity field is described by a Lindblad type master equation  (see Supplementary Note 1 for derivation)
\begin{equation}
\dot{\rho}(t)=-\frac{i}{\hbar}[H,\rho(t)]+\Gamma\_r\bar{n}\_{th} \mathcal{L}[a^\dagger]\rho(t)+\Gamma\_r(\bar{n}\_{th}+1) \mathcal{L}[a]\rho(t),
\label{eq:3}
\end{equation}
where  $\rho(t)$ is the cavity field density matrix and the Lindblad operator is defined as $\mathcal{L}[a]\rho=a\rho a^\dagger -\frac{1}{2}\left(a^\dagger a \rho + \rho a^\dagger a \right)$ pertaining to the annihilation operator $a$.
The Hamiltonian is given by
\begin{equation}
H/\hbar=\gamma\_{inj}\sinc(\Delta\_{ac}\tau)g\tau(\rho\_{eg}a^\dagger e^{-i\Delta\_{ac}\tau/2}+\rho\_{ge}a e^{i\Delta\_{ac}\tau/2}),
\label{eq:hamiltonian}
\end{equation}
where $\gamma\_{inj}\equiv \bar{N}/\tau$ denotes the atomic injection rate, $\tau$ is the interaction time, $\bar{N}$ is the mean number of atoms in the cavity, $\Delta\_{ac}$ is the atom-cavity detuning, and $\rho\_{eg(ge)}$ is the off-diagonal element of the density matrix of the atom. The thermal photon number $\bar{n}\_{th}$ and the decay constant $\Gamma\_r$ are given by $\bar{n}\_{th}\equiv\frac{\rho\_{ee}\gamma\_{inj}  (g\tau)^2 \sinc^2(\Delta\_{ac}\tau/2) }{(\rho\_{gg}-\rho\_{ee}) \gamma\_{inj} (g\tau)^2 \sinc^2(\Delta\_{ac}\tau/2)  + 2\kappa}$ and $\Gamma\_r\equiv (\rho\_{gg}-\rho\_{ee}) \gamma\_{inj} (g\tau)^2 \sinc^2(\Delta\_{ac}\tau/2) + 2\kappa$, respectively.

The second (third) term in Eq.~(\ref{eq:3}) containing Lindblad operator $\mathcal{L}[a^\dagger](\mathcal{L}[a] )$ include both the effects of atoms and photonic vacuum, showing that
both can be grouped together as if the cavity were connected to a single heat reservoir. The second(third) term corresponds to the upper(lower) state of the reservoir.
Therefore, the ratio of the coefficients can be associated with the effective reservoir temperature $T\_R$ through a Boltzmann factor, 
$\frac{\bar{n}\_{th}}{\bar{n}\_{th}+1}=\exp\left(-\frac{\hbar\omega\_a}{k\_B T\_R} \right)$,
and thus
\begin{equation}
\begin{split}
T\_R&=\frac{\hbar \omega\_a}{k\_B\log \left(1+\bar{n}\_{th}^{-1}\right)}\\
&=\hbar \omega\_a\left\{k\_B\log \left[ \frac{\rho\_{gg}}{\rho\_{ee}}+\frac{2\kappa}{\gamma\_{inj}\rho\_{ee}(g\tau)^2 \sinc^2(\Delta\_{ac}\tau/2)} \right]\right\}^{-1}.
\end{split}
\label{eq:tr}
\end{equation}

Equation (\ref{eq:tr}) is reduced to the Boltzmann factor $\rho\_{ee}/\rho\_{gg} = e^{-\hbar \omega_a / k\_B T\_R}$ when the cavity is lossless($\kappa=0$).
The Hamiltonian in Eq.~(\ref{eq:hamiltonian}) contains a cavity driving term proportional to the off-diagonal element of atomic state. This term describes superradiance originating from the coherence injection.\\

\noindent
{\bf Effective temperature of photons.}
During the isoenergetic expansion and compression processes, the effective temperature of the photons can be deduced by the thermodynamic relation $dQ=TdS$.
First, the entropy can be written as
\begin{equation}
S/k\_B=( {n}\_{th}+1)\log( {n}\_{th}+1)- {n}\_{th}\log \left( {n}\_{th}\right),
\label{eq:13}
\end{equation}
for both the thermal and thermal coherent states (see Supplementary Note 3 for derivation).
Differentiation of Eq.~(\ref{eq:13}) gives
\begin{equation}
\begin{split}
\frac{dS}{k\_Bd {n}\_{th}} &= \log \left(1+ {n}\_{th}^{-1}\right) .
\end{split}
\end{equation}
For the isoenergetic processes, the infinitesimal change in heat can be written as $dQ=dW=-\sum_k{P_kdE_k}=-\hbar \sum_k{kP_kd\omega_k}=-n\hbar d\omega\_c$, so the effective engine or cavity temperature  $T\_c$ is 
\begin{equation}
\begin{split}
T\_c&=dQ/dS=-\frac{n\hbar d\omega\_c}{k\_B\log \left(1+ {n}\_{th}^{-1}\right)d {n}\_{th}}\\
&=\frac{n\hbar \omega\_c}{k\_B\log \left(1+ {n}\_{th}^{-1}\right)  {n}\_{th}},
\end{split}
\label{eq:15}
\end{equation}
where the internal energy conservation, $dU=d {n}\_{th}\hbar\omega\_c+ {n}\_{th}\hbar d\omega\_c=0$, was used.

The electromagnetic field in a cavity can be treated as a one-dimensional quantum harmonic oscillator, which has two degrees of freedom. According to the equipartition theorem, the internal energy of the photons $U(=n\hbar \omega\_c)$ should thus be equal to $k\_BT$ in the thermodynamic limit.
This is confirmed in Eq.~(\ref{eq:15}): as $ {n}\_{th}$ diverges, $\log \left( 1+{n}\_{th}^{-1}\right)  {n}\_{th}$ approaches unity, thus Eq.~(\ref{eq:15}) is reduced to $n\hbar \omega\_c=k\_BT$. \\

\noindent
{\bf Efficiency of superradiant quantum heat engine.}
Our engine cycle consist of isochoric and isoenergetic processes.
Heat and work of each stroke can be calculated by the quantum mechanical version of the first law of thermodynamics\cite{quan2007quantum}.
The symbols related to the photon number ($n\_{sr},n'\_{sr},n\_{th},n'\_{th}$) and the cavity resonance frequency ($\omega\_{c1},\omega\_{c2}$) are illustrated somewhat exaggeratedly in Extended Data Fig.~\ref{extfig:1}.
In the isochoric heating process A$\rightarrow$B, the internal energy of the engine is increased by
\begin{equation}
\begin{split}
\Delta E\_{AB}&=\sum_k{\int_A^B{E_k dP_k}} \\
&=\sum_k{k\hbar\omega\_{c1}\int_A^B{ dP_k}}= (n\_{sr}-n\_{th}) \hbar \omega\_{c1},
\end{split}
\end{equation}
where $E_k(=k\hbar\omega\_c)$ and $P_k$ is the eigenenergy and probability of the $k$th eigenstate, respectively.
Since the entropy of the engine is unchanged during this process, the energy gain cannot be explained by the classical thermodynamics. This energy change originates from the ergotropy $\mathcal{E}$ of the non-passive superradiant state of the reservoir. The ergotropy transfer, equal to $\Delta E\_{AB}=-\Delta\mathcal{E}\_{AB}(>0)$, is sometimes called `internal work' because it occurs without  entropy change of the engine\cite{niedenzu2016operation}. 
Alternatively, the ergotropy transfer can be interpreted as heat $Q\_{AB}$ coming from the change $\Delta \mathscr{C}\_{AB}$  of the relative entropy of coherence of the non-passive reservoir as $Q\_{AB}=-\Delta \mathcal{E}\_{AB}=-T\_R \Delta \mathscr{C}\_{AB}$ as shown in Supplementary Note 4.

During the isoenergetic expansion B$\rightarrow$C, the work done by the engine is given as
\begin{equation}
\begin{split}
W\_{BC}&=-\sum_k{\int_B^C{P_k dE_k}}=-\int_B^C{\hbar\sum_k{k P_k d\omega\_c}} \\
&=-\int_B^C{n \hbar \frac{\omega\_{c}}{\omega\_{c}}d\omega\_c}=-\int_B^C{\frac{n\_{sr} \hbar \omega\_{c1}}{ \omega\_{c}} d\omega\_c} \\
&= n\_{sr} \hbar \omega\_{c1}\log{\frac{\omega\_{c1}}{\omega\_{c2}}},
\end{split}
\label{eq:8}
\end{equation}
where the internal energy conservation $n'\_{th}\hbar\omega\_{c2}=n\_{th}\hbar\omega\_{c1}$ and $n'\_{sr}\hbar\omega\_{c2}=n\_{sr}\hbar\omega\_{c1}$ is taken into account. 
According to the first law of thermodynamics, the work $W\_{BC}$ equals the absorbed heat  $Q\_{BC}$ from the reservoir in the isoenergetic process,
\begin{equation}
Q\_{BC}=W\_{BC}=n\_{sr} \hbar \omega\_{c1}\log{\frac{\omega\_{c1}}{\omega\_{c2}}},
\end{equation}
which can be equated with $T\_c \Delta S\_{BC}$ using Eq.~(\ref{eq:15}). We can interpret the heat $Q\_{BC}$ in terms of the relative entropy of coherence of the reservoir as $Q\_{BC}=T\_R \Delta S\_{BC}-T\_R\Delta \mathscr{C}\_{BC}$, where the second term is identified as the ergotropy transfer $-\Delta \mathcal{E}\_{BC}= -T\_R \Delta \mathscr{C}\_{BC}(>0)$ from the superradiant reservoir. 

Similarly, we obtain the expressions for heat, work and ergotropy transfer in the other processes as
\begin{equation}
\begin{split}
Q\_{CD}&=-\Delta \mathcal{E}\_{CD}= (n'\_{th}-n'\_{sr}) \hbar \omega\_{c2},=-Q\_{AB} \\
W\_{DA}&=-n'\_{th} \hbar \omega\_{c2}\log{\frac{\omega\_{c1}}{\omega\_{c2}}}, \\
Q\_{DA}&=W\_{DA}=T\_R \Delta S\_{DA}=- n'\_{th} \hbar \omega\_{c2}\log{\frac{\omega\_{c1}}{\omega\_{c2}}}=-Q\_{out}.
\end{split}
\end{equation}
Note the heat $Q\_{AB}$ or ergotropy transfer $-\Delta \mathcal{E}\_{AB}$ from the reservoir to the engine during the isochoric process A$\rightarrow$B is returned to the reservoir in the other isochoric process C$\rightarrow$D, {\em i.e.} $Q\_{CD}=-Q\_{AB}$.  The net heat absorbed and the net work done by the engine per cycle are thus described by
\begin{equation}
\begin{split}
Q\_{in}&=Q\_{AB}+Q\_{BC}+Q\_{CD}=Q\_{BC}=n\_{sr} \hbar \omega\_{c1}\log{\frac{\omega\_{c1}}{\omega\_{c2}}}, \\
W\_{out}&=W\_{BC}+W\_{DA}=(n\_{sr}-n\_{th}) \hbar \omega\_{c1}\log{\frac{\omega\_{c1}}{\omega\_{c2}}} \\
&=Q\_{in}-Q\_{out}.
\end{split}
\end{equation}
The efficiency, defined as the ratio $W\_{out}/Q\_{in}$, is then given by the ratio of the number of photons as
\begin{equation}
\eta=W\_{out}/Q\_{in}=1-\frac{n\_{th}}{n\_{sr}}.\\
\label{eq:14}
\end{equation}
Note that the net work can also be written as  $W\_{out}=T\_R\Delta S\_{BC}-\Delta \mathcal{E}\_{BC}+T\_R\Delta S\_{DA}=-\Delta \mathcal{E}\_{BC}$ since $\Delta S\_{BC}=-\Delta S\_{DA}$, clearly showing that it entirely comes from the ergotropy transfer from the superradiant reservoir occurring in the expansion stage. \\

\hbox{}
\hbox{}

\noindent
{\bf \large Data availability}

\noindent
The datasets generated during the current study are available from the corresponding author on reasonable request.\\

\noindent
{\bf \large Code availability}

\noindent
The code that supports the findings of this study are available from the corresponding author upon reasonable request.\\

\hbox{}
\hbox{}

\noindent\textbf{Acknowledgements}

\noindent
This work was supported by the Korea Research Foundation(Grant No. 2020R1A2C3009299) and the Ministry of Science and ICT of Korea under ITRC program (Grand No. IITP-2021-2018-0-01402).\\

\noindent\textbf{Author Contributions}

\noindent
Ji.K. and K.A. conceived the experiment. Ji.K performed the experiment with help from S.O. and analyzed the data and carried out theoretical investigations. K.A. supervised overall experimental and theoretical works. Ji.K. and K.A. wrote the manuscript. All authors participated in analyses and discussions.\\

\noindent\textbf{Competing Interests}

\noindent
The authors declare no competing interests.\\

\begin{extfigure}
\centering
\includegraphics[width=0.6\textwidth]{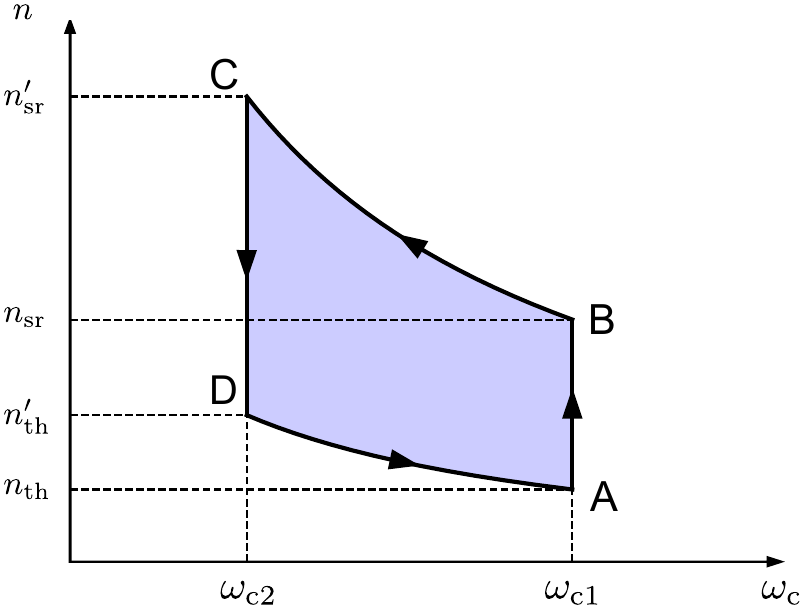}
\caption{{\bf The pressure-volume diagram.} The pressure-volume diagram of the superradiant quantum engine. The radiation pressure is represented by the photon number. The $x$ axis is the cavity resonance frequency, which is inversely proportional to the cavity mode volume.
}
\label{extfig:1}
\end{extfigure}

\end{document}